\documentclass[preprint,3p,times,authoryear,twocolumn]{elsarticle}

\usepackage{amssymb}
\usepackage{amsthm}
\usepackage{amsmath}
\usepackage{scalefnt}
\usepackage{hyperref}
\usepackage{scalefnt}
\usepackage{multirow}
\usepackage{setspace}
\usepackage{xcolor}
\usepackage{soul}\usepackage{graphicx}
\usepackage{dcolumn}
\usepackage{bm}
\usepackage{xcolor}
\usepackage[bf]{subfigure}

\journal{Social Networks}

\begin{document}

\begin{frontmatter}



\title{Opinion Diversity and Social Bubbles in Adaptive Sznajd Networks}

\author[label1]{Alexandre Benatti}
\author[label1]{Henrique F. de Arruda}
\author[label2]{Filipi Nascimento Silva}
\author[label3]{C\'esar Henrique Comin}
\author[label1]{Luciano da Fontoura Costa}

\address[label1]{S\~ao Carlos Institute of Physics, University of S\~ao Paulo, S\~ao Carlos, SP, Brazil.}
\address[label2]{School of Informatics, Computing and Engineering, Indiana University, Bloomington, Indiana 47408, USA.}
\address[label3]{Department of Computer Science, Federal University of S\~ao Carlos, S\~ao Carlos, Brazil}


\begin{abstract}
Among the several approaches that have been attempted at studying opinion dynamics, the Sznajd model provides some particularly interesting features, such as its simplicity and ability to represent some of the mechanisms believed to be involved in opinion dynamics.  The standard Sznajd model at zero temperature is characterized by converging to one stable state, implying null diversity of opinions.  In the present work, we develop an approach -- namely the adaptive Sznajd model -- in which changes of opinion by an individual (i.e. a network node) implies in possible alterations in the network topology. This is accomplished by allowing agents to change their connections preferentially to other neighbors with the same state.  The diversity of opinions along time is quantified in terms of the exponential of the entropy of the opinions density.  Several interesting results are reported, including the possible formation of echo chambers or social bubbles.  Additionally, depending on the parameters configuration, the dynamics may converge to different equilibrium states for the same parameter setting, which suggests that this phenomenon can be a phase transition. The average degree of the network strongly influences the resultant opinion distribution, which means that echo chambers are easily formed in lower connected systems. 
\end{abstract}


\end{frontmatter}


\section{\label{sec1}Introduction}
An important property of complex systems involving several components regards the distribution of their components along time. For instance, consider the distribution of the accumulated charge in a biological neuronal networks~\citep{dalla2019modeling}.  When many neurons are near the firing threshold, even relatively minute events can imply avalanches of activation.  Similar situations can be found in economy, scientific development, ecology, arts, among many other areas.  Interestingly, many of these systems can be represented in terms of a respective complex network.

Therefore, it becomes interesting not only to quantify the diversity of individuals in given complex systems along time, but also to study how these systems tend to yield more or less uniform state distributions.  In addition, it is also important to understand how modifications in such systems, e.g.,~by increasing or reducing inter-connectivity, can influence the respective diversity of states.  

Understanding the way in which human opinion changes along time and space constitutes one of the great challenges in complex systems research. The importance of such studies has been reflected in the proposal of several modeling approaches, such as the Sznajd model \citep{sznajd2000opinion}, which is an approach to opinion dynamics. This model is particularly interesting because it takes into consideration that two or more friends sharing the same opinion are more likely to persuade other friends. In other words, a person is more susceptible to change opinion $o$ when many of their friends have that same opinion $o$.
The complex system of interest can be represented as a network, and topological interactions can be made to influence the dynamics of opinions.  More specifically, it is frequently adopted that when a pair of neighboring nodes share the same opinion $o$, the nodes connected to this pair tend to be influenced by $o$.  This dynamic depends on a parameter of temperature, $w$, that controls the spontaneous change of opinion~\citep{araujo2015mean}. 

The characterization of the diversity of the opinion dynamics has often been approached in terms of relative frequencies of the opinions, while the consideration of information theory principles such as entropy are more rarely employed. For instance, in~\citep{messias2018can} the authors investigated the relationship between the rewiring parameter of the Watts-Strogatz model and the diversity of opinions using the Sznajd model. They found that opinion diversity tends to be inversely proportional to the amount of shortcuts in the network. For quantifying the diversity of opinions, they calculated the measurement based on the entropy of the distribution of opinions in the network~\citep{jost2006entropy}. Interestingly, the time required for the dynamics to reach the steady-state was found to be an essential factor for quantifying diversity in the Sznajd dynamics.

Because low diversity can be undesired in some situations, it becomes interesting to devise means to promote opinion diversity. The current work approaches this issue by considering the Sznajd dynamics and remodeling of the network topology by changing the opinion of some nodes and reassigning their connectivity in the network.  More specifically, if a node changes opinion from $o_1$ to $o_2$, it can be disconnected from its neighbours having opinion $o_1$ and reconnected, with probability $q$, to another node having opinion $o_2$. Other studies have also incorporated a rewiring dynamics~\citep{he2004sznajd,holme2006nonequilibrium,fu2008coevolutionary,durrett2012graph,iniguez2009opinion}.

In social networks, people usually have many connections, but it is expected that the majority of them can communicate only with a small number of friends because of time constraints. Furthermore, many of the social networking platform automatically suggest connections between individuals who share common interests. Due to these characteristics, we propose a dynamics that is executed in a network with a limited number of connections and with the possibility of opinion induced rewiring. This type of network modification is hoped to promote the diversity of opinions in the simulated system.  However, several parameters can influence this effect, including the temperature $w$, the network average degree $\langle k \rangle$, the rewiring probability $p$, and the opinion induced rewiring probability $q$. We perform several experiments in order to identify the effect, isolatedly or combined, of these parameters on the diversity of opinions. Interestingly, this dynamics can result in a network with more than one connected component, which can be associated with social bubbles.

Several results are reported and discussed, including the capability of the proposed dynamics to simulate situations that result in social bubbles, the fact that the computed diversity can be strongly affected by the probability of an agent spontaneously changing its opinion, and, the possible emergence of two distinct types of opinion distributions with the same dynamics configuration. As in the standard Sznajd model, the increase in the temperature led to faster convergence times. Additionally, in the proposed dynamics, the computed diversity can be strongly affected by the temperature. Depending on the employed parameters, our dynamics results in an ``echo chamber'' (also known as ``filter bubbles''), in which the agents can be connected only with others that have the same opinion.  This result agrees with studies performed on real social networks~\citep{nikolov2015measuring}.

The remaining of this paper is organized as follows. In section~\ref{sec:related}, we briefly describe some previous approaches to opinion state dynamics. In section~\ref{sec2}, we present the Sznajd model. In section~\ref{sec:met}, we describe the employed methodology as well as the diversity measurement and the used network model. In section~\ref{sec:results}, we present the results and discussions. Finally, in section~\ref{sec:conc}, we provide the conclusions.

\section{\label{sec:related}Previous approaches}
Many previous works have investigated the Sznajd model~\citep{castellano2009statistical}. For example,  in~\citep{he2004sznajd} the authors include a parameter of social temperature (the probability of an agent accepting another agent's opinion). In~\citep{gonzalez2004opinion}, the Sznajd model is used for modeling the distribution of votes in an election process. Some models incorporating a continuous opinion formulation have also been defined~\citep{fortunato2005sznajd}. Furthermore, there are many other proposed approaches for modeling opinion dynamics~\citep{castellano2009statistical, dong2018survey}. For example, in~\citep{gomes2019mobility} the authors describe a model where the agents explore other opinions in a time-varying network using a random walk dynamics.

An approach based on the voter model is proposed in~\citep{holme2006nonequilibrium}, where a node $i$ is selected at random, and with a given probability, the agent $i$ can rewire one edge and connect to a target node that has the same opinion as $i$. This mechanism is similar to the proposal for this study. However, in the current work we employ a more elaborate opinion dynamics that incorporates the characteristics of the Sznajd model and the rewires are not chosen at random.

Mean-field analysis has also been employed to a variation of the voter model, where~\cite{vazquez2008generic} found that there is an absorbing transition from two coexisting states to a single state. In a different approach~\citep{fu2008coevolutionary}, the authors also propose a rewiring dynamics which aims at avoiding connections with minority and different opinions, in which the new target connections are set to neighbors of neighbors.

Other approaches include the dynamics of cultural formation, e.g., the well-known Axelrod~\citep{axelrod1997dissemination} model. Such an approach often considers the complex system of interest to be represented as a complex network. In the Axelrod approach, there is a vector of $F$ features taking discrete values. These features are allowed to change so that two neighboring nodes can influence one another, resulting in more similar vectors. Other methods have been derived from the Axelrod model, such as~\citep{gracia2009residential}, which takes into account the movement of the agents, and~\citep{rodriguez2010effects}, which considers the influence of mass media on the opinion dynamics. Some related approaches also consider co-evolving networks, which are network topologies that vary along time, e.g., cultural formation~\citep{centola2007homophily} and  socio-economical dynamics~\citep{biely2009socio}.

\section{\label{sec2}The Sznajd model}
Many different dynamics have been proposed to simulate the transmission of opinions~\citep{hegselmann2002opinion}. Furthermore, some of these approaches take into consideration the structure of complex networks, such as in~\citep{rodrigues2005surviving, gonzalez2006renormalizing, bernardes2002election}. In this study, we focus on the Sznajd model~\citep{sznajd2000opinion}.

Because there are some variations of the Sznajd dynamics on networks, we considered the version adopted in~\citep{bernardes2002election}. In this model, the network nodes are the agents, $N_O$ defines the maximum number of opinions, and the agents can also be in an undecided state (null opinion).
The undecided state contributes to the realism of the simulation since, in real situations, when a new subject arises many people would not yet have formed their respective opinions.
The dynamics start with agents having opinions randomly chosen with uniform probability. For each iteration, a node $i$ is selected, and the opinions can change according to the dynamics rules, as follows:

\begin{itemize}
\item If the node $i$ is undecided nothing happens; 
\item If the node $i$ has an opinion $o$, one of the $i$ neighbors, $j$, is randomly selected with uniform probability. By considering the agents $i$ and $j$, the following rules are applied:
  \begin{itemize}
    \item If $j$ is assigned as undecided, the opinion of $j$ is assigned as $o$ with probability inversely proportional to the $i$ degree;
    \item if $i$ and $j$ have the same opinion, than their neighbors are assigned as $o$ with probability inversely proportional to the $i$ and $j$ degrees, respectively;
    \item If $i$ and $j$ have different opinions, nothing happens.
  \end{itemize}
\end{itemize}

Apart from these rules, there is another parameter, $w$ ($0 \leq w \leq 1$ ), called temperature, which is used to control the probability of an agent to change its opinion randomly, with uniform probability among the remaining opinions. This change does not assign the node to the undecided opinion. 

\section{\label{sec:met} Methodology}
In this section we describe the adopted variation of the Sznajd model and the employed methodology to quantify the results, including the definition of the diversity measurement~\citep{jost2006entropy}.

\subsection{Adaptive Sznajd Model}

We adopt the Sznajd model to simulate friendship dynamics, which consequently can promote diversity of opinions and can lead to social bubbles. For that, we included rules that control the rewiring of the edges, which modify the structure of the network according to the agent's opinions. This dynamics is henceforth called \emph{Adaptive Sznajd Model} (ASM). At the same iteration of the Sznajd model, if one or more nodes change their opinions, the following rules are applied with probability $q$ (opinion induced rewiring probability):
\begin{itemize}
\item If an agent changes its opinion to another that is different from the opinions of all other nodes, nothing happens;
\item If the new opinion is equal to all of its neighbors, nothing happens;
\item If the two above situations did not occur, the node deletes one connection with a node that has a different opinion and connects to another node having the same opinion. This connection is randomly selected with uniform probability. 
\end{itemize}
The above rules are not employed when the agent's opinion changes only because of the temperature.
Table~\ref{tab:parameters} presents a social interpretation of the parameters involved in the currently adopted model.

\begin{table}[h]
\centering
\caption{Social interpretation of the used parameters.}
\label{tab:parameters}
\vspace{0.4cm}
\small
\begin{tabular}{|c|c|}
\hline 
\textbf{Dynamics Parameter} & \textbf{Social Interpretation} \\
\hline 
\hline 
\multirow{2}*{Temperature ($w$)} & \multicolumn{1}{p{3.3cm}|}{Susceptibility to change the opinion spontaneously} \\
\hline 
Rewiring Probability ($q$) & \multicolumn{1}{p{3.3cm}|}{Friendship volatility} \\
\hline 
WS rewiring probability ($p$) & \multicolumn{1}{p{3.3cm}|}{Spatiality of friendships} \\
\hline
Average Degree ($\langle k \rangle$) & \multicolumn{1}{p{3.3cm}|}{Social cohesion} \\  
\hline 
\end{tabular}
\label{table1}
\end{table}

\subsection{Diversity}
In order to quantify the effective number of states, many researchers from distinct areas have been employing measurements of diversity~\citep{jost2006entropy}. For example, to compute diversity of species in ecology~\citep{leinster2012measuring}, to calculate properties of networks~\citep{travenccolo2008accessibility}, to identify influential spreaders in information diffusion~\citep{de2014role}, and to analyze brain regions~\citep{betzel2018diversity}. Additionally, in~\citep{messias2018can} the authors quantified the diversity of opinions in the Sznajd model. As in~\citep{messias2018can}, in order to compute diversity, here we considered each opinion as representing one different state.

The diversity is defined based on information theory, more specifically on the Shannon entropy of the opinions frequency in the network. The diversity is calculated as
\begin{equation}
    D = \exp{(H)},
\end{equation}
where $H$ is the the Shannon entropy, which is defined as
\begin{equation}
    H = -\sum_{o=1}^n{\rho_o \ln(\rho_o)},
\end{equation}
where $\rho_o$ is the relative frequency of state $o$ and $n$ is the number of states (number of different opinions). Here, $n = N_O + 1$, because of the null opinion. Taking an example in which $\rho_1 = 1$ and the remaining states are zero, then $D = 1$. Furthermore, in the case when $\rho_1 = 0.5$, $\rho_2 = 0.5$, and the remaining states are zero, $D = 2$. In other words, if there is a single state, $D = 1$, and if there are $x$ states with similar probabilities, $D \approx x$. In summary, the measurement of diversity $D$ expresses the effective number of opinions in the network.

\subsection{The adopted network model}
In this study, we test our dynamics on Watts-Strogatz (WS)  networks~\citep{watts1998collective}. This model starts with a regular network with $N$ nodes, here we adopt a regular 2D lattice having toroidal boundary condition. Each edge of the network is rewired with probability $p$. If a given edge $(i,j)$ is to be rewired, a new node $l\neq i$ is randomly chosen with uniform probability, and the edge is altered to $(i,l)$. Here, we fixed the number of nodes as $N=1089$. We chose this network model because it allows the generation of networks ranging from a lattice-like topology ($p\approx 0$) to a completely random connectivity ($p=1$).

\subsection{Identifying the equilibrium states}

\begin{figure*}[t]
  \centering
    \subfigure[$\langle k \rangle = 8$, $q=0$, $p=0.001$, and $w=0$.]{\includegraphics[width=0.48\textwidth]{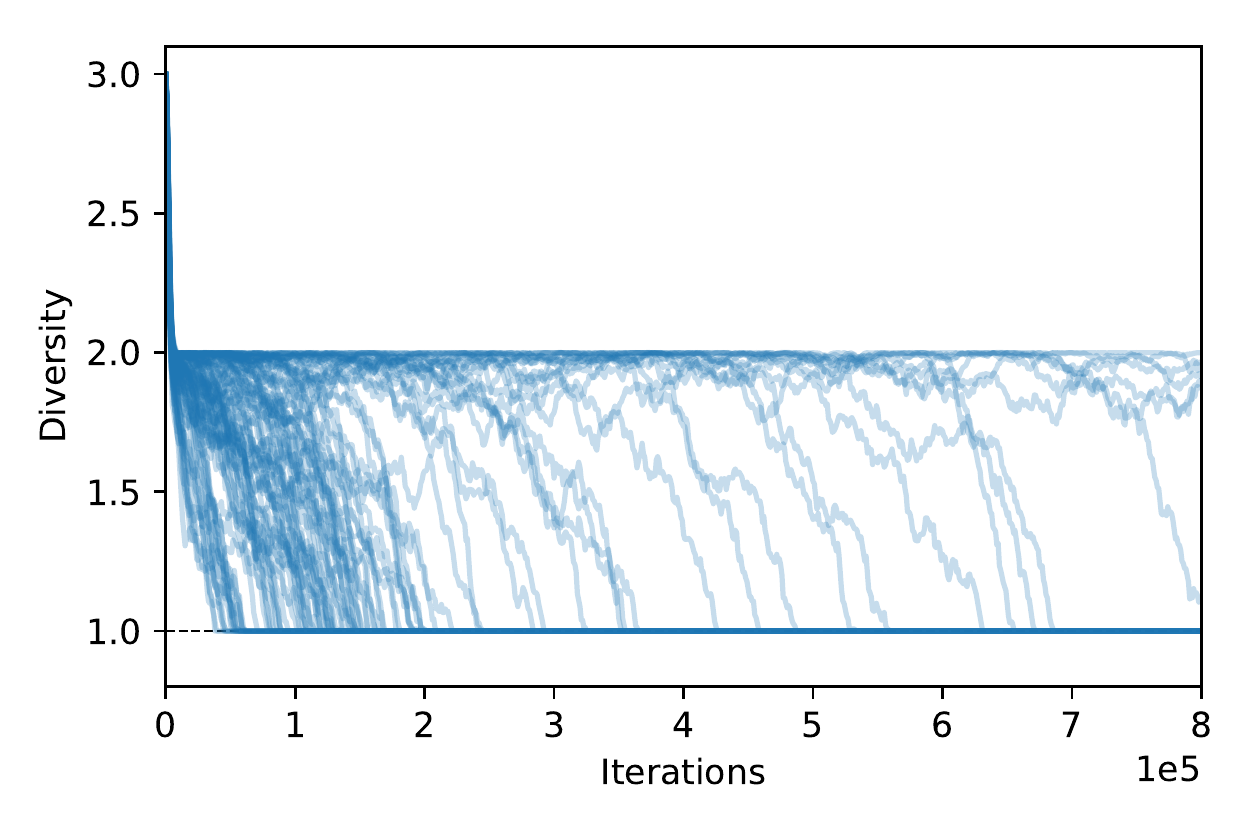}}
    \subfigure[$\langle k \rangle = 8$, $q=1$, $p=0.001$, and $w=0$.]{\includegraphics[width=0.48\textwidth]{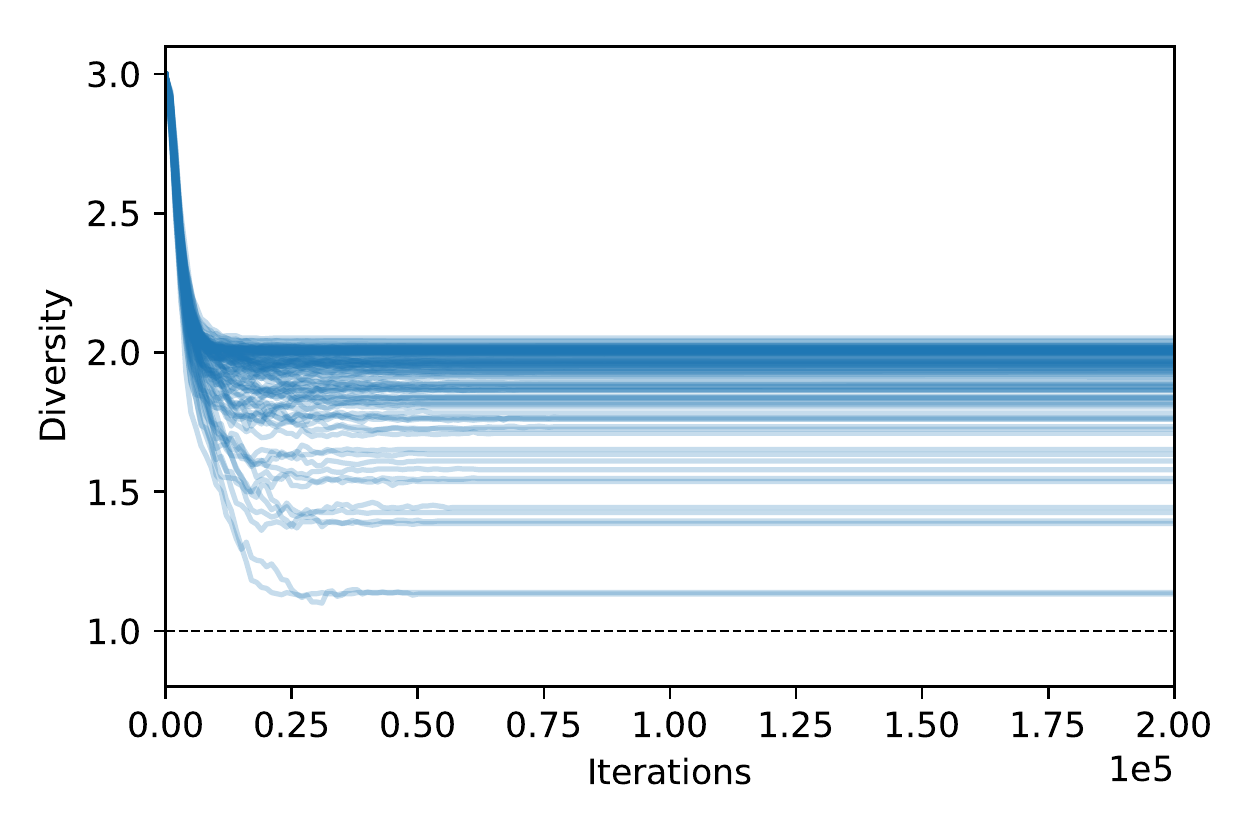}}
    \subfigure[$\langle k \rangle = 8$, $q=0.6$, $p=0.001$, and $w=0$.]{\includegraphics[width=0.48\textwidth]{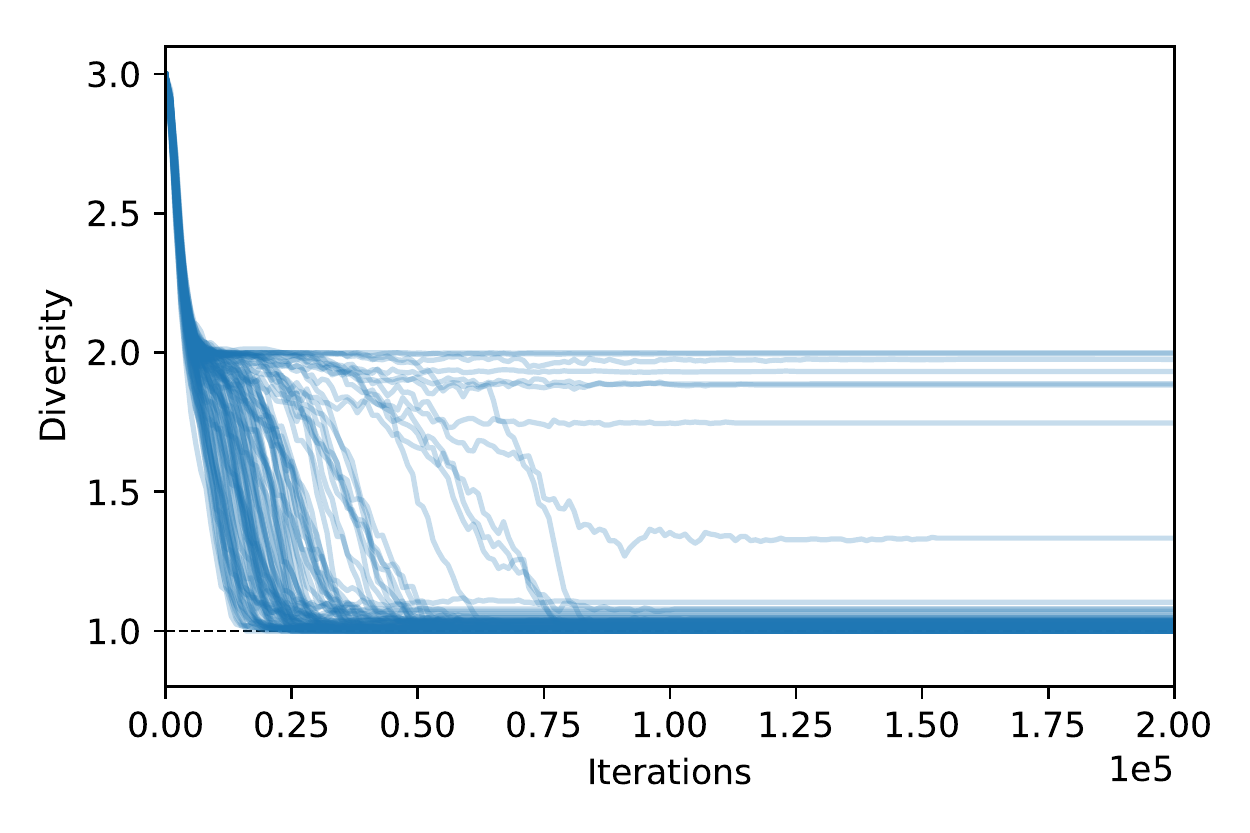}}
    \subfigure[$\langle k \rangle = 8$, $q=0.6$, $p=0.001$, and $w=0.05$.]{\includegraphics[width=0.48\textwidth]{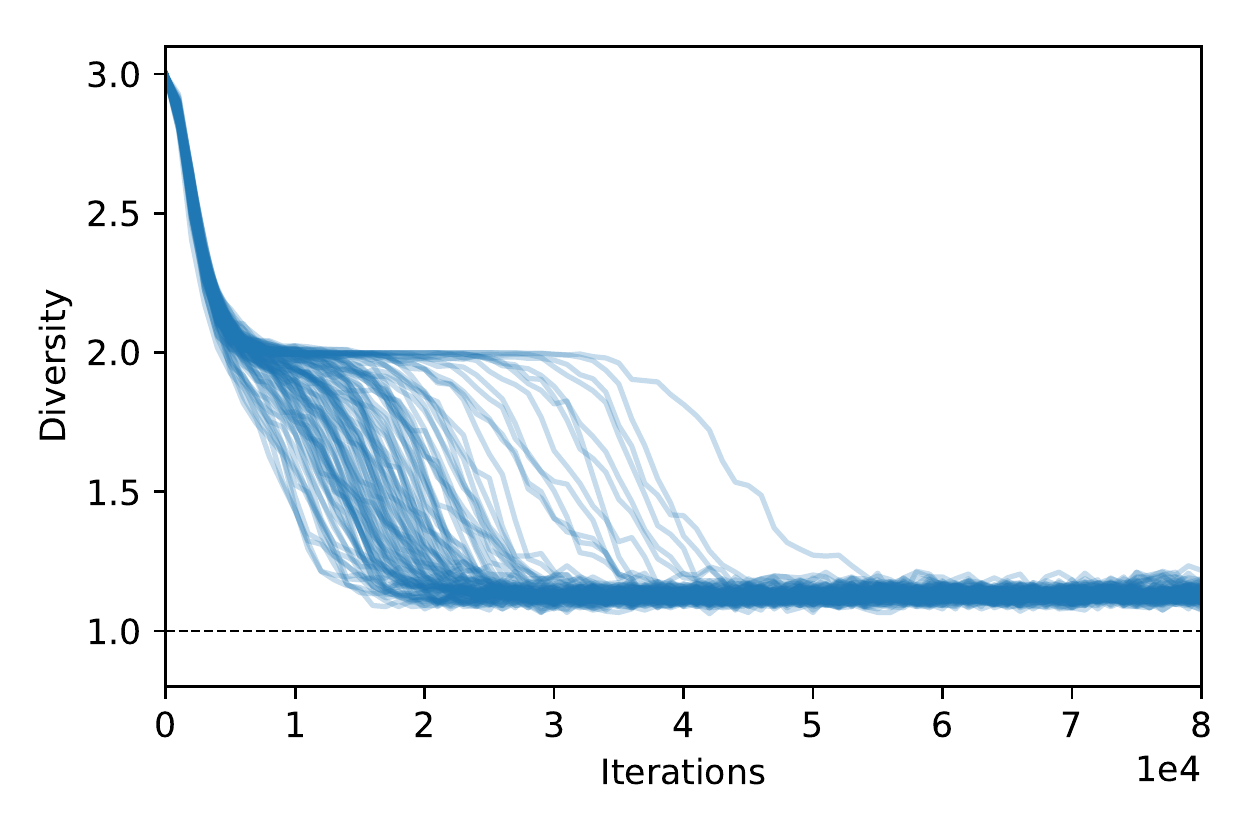}}
  \caption{Diversity of opinions as a function of the number of iterations of the dynamics. The parameters used in each experiment are indicated below each plot. Each curve represents a different executions of the dynamics.}
  \label{fig:diversityXtime}
\end{figure*}

Since the considered dynamics will exhibit, for each given network topology, varying diversity along time, it is important to devise means to standardize the time interval in which the diversity is studied.  Figure~\ref{fig:diversityXtime} shows examples of the unfolding of diversity along time for four different dynamics configuration, namely: (a) $\langle k \rangle = 8$, $q=0$, $p=0.001$, and $w=0$ (this is the standard Sznajd model); (b) $\langle k \rangle = 8$, $q=1$, $p=0.001$, and $w=0$; (c) $\langle k \rangle = 8$, $q=0.6$, $p=0.001$, and $w=0$; and (d) $\langle k \rangle = 8$, $q=0.6$, $p=0.001$, and $w=0.05$.  Observe that temperature is set to zero for cases (a) to (c), while case (d) considers a non-zero temperature.

An interesting result, also found for many other configurations, is that the diversity always tend to an equilibrium value, after which it remains constant or nearly so (in the case of non-zero temperature).  So, it is possible to perform our analysis of diversity after equilibrium is achieved, which is henceforth adopted. For each set of parameter values used in the experiments, the number of iterations necessary for achieving the equilibrium state was visually determined.

\section{\label{sec:results} Results and discussion}

In the following, we present the results and their respective discussion regarding the proposed Sznajd model, describing the influence of temperature ($w$), opinion induced rewiring probability ($q$) and average network degree ($\langle k \rangle$) on the diversity of opinions.

\subsection{Convergence in the ASM}
We begin by analyzing the diversity of opinions as a function of the number of iterations of the dynamics, shown in Figure~\ref{fig:diversityXtime}. Each plot shows 100 executions of the dynamics using the same network parameters. In the case of Figure~\ref{fig:diversityXtime}(a), which was computed with $q=0$ (the standard Sznajd model), the diversity always goes to 1, but for some cases it may take a long time to reach this consensus state. In contrast with this result, by considering $q=1$, (see Figure~\ref{fig:diversityXtime}(b)), the dynamics leads to a diversity higher than one. So, there is more than one opinion in the final state of the dynamics. Furthermore, in most cases the obtained diversity is close to 2, which indicates that each opinion is held by half of the nodes. The cases with diversity higher than 2 happen due to the presence of isolated nodes having the null opinion. By considering an intermediate value of the opinion induced rewiring probability ($q=0.6$), in the majority of the cases, the dynamics leads the diversity to be near 1 (see Figure~\ref{fig:diversityXtime}). However, there is a possibility to have diversity equals to 2. By employing the same parameters, but also including temperature ($w=0.05$), the measured diversity tends to values near one.  In addition, the transient time reduces significantly (one order of magnitude).

\subsection{Parameters analysis of the ASM}

\begin{figure}[!htpb]
  \centering
     \includegraphics[width=0.48\textwidth]{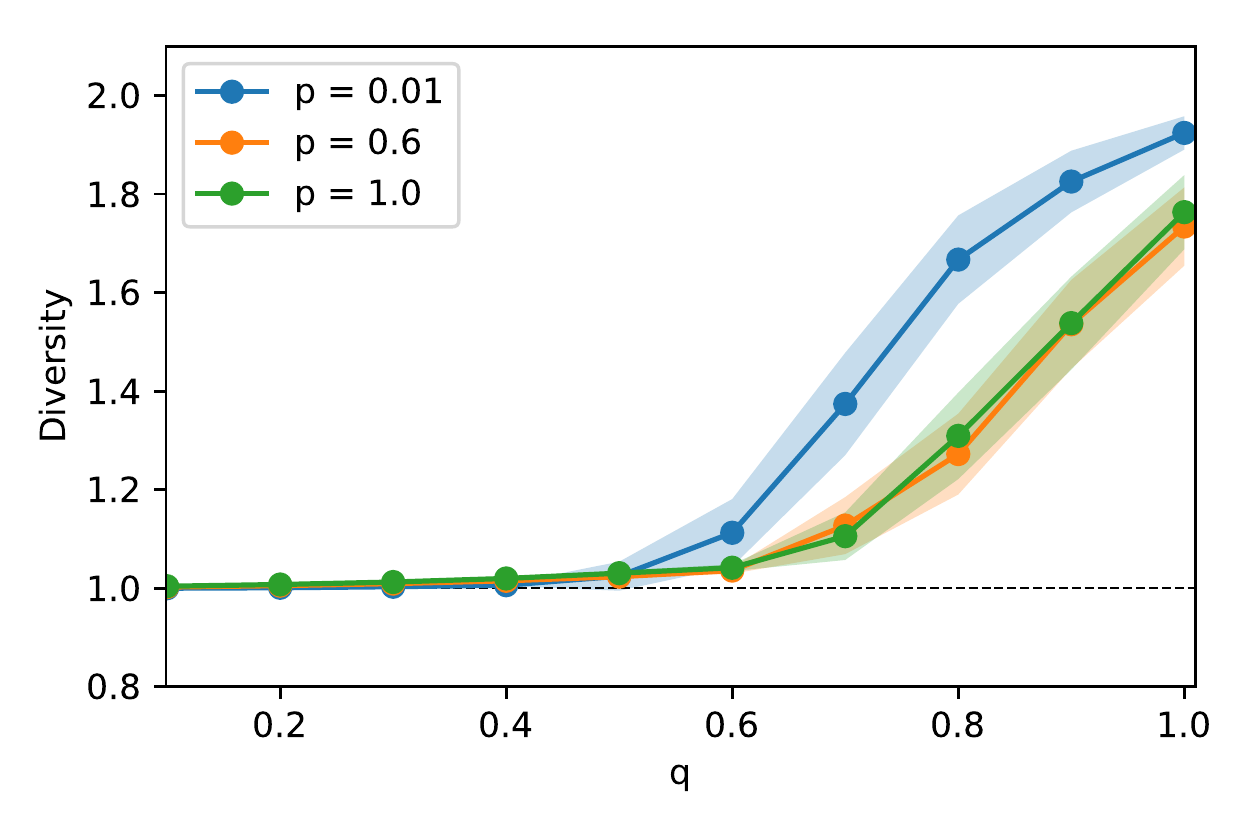}
   \caption{Average diversity of opinion as a function of the opinion induced rewiring probability ($q$) for different rewiring probabilities ($p$). The averages were calculated over 100 realizations of the dynamics. Shaded regions represent $25\%$ of the standard deviation respectively to the curves with the same color.}
  \label{fig:diversityXq}
\end{figure}

In order to better understand how diversity changes according to the initial configuration of the networks and parameter $q$ of the dynamics, we plot the average curves of diversity and the respective standard deviations against $q$ for distinct values of rewiring parameter $p$. Here we considered some selected sets of parameters, as shown in Figure~\ref{fig:diversityXq}. The high values of standard deviations are found because the diversity can attain many different values at the equilibrium. This effect is observed for many of the tested parameters, as shown in Figure~\ref{fig:diversityXtime}. So, in order to better visualize the results, we plot only $25\%$ of the standard deviations. All values were obtained after the dynamics reached the equilibrium, and the dynamics was executed 100 times for each set of parameters. The results indicate that lower values of $p$ result in higher diversity. In other words, the more organized the network is, the more diverse the opinions become. Furthermore, for $q<0.5$, the average diversity tends to be close to one, while for $q>0.5$ the diversity increases according to $q$.  

\begin{figure}[!htpb]
  \centering
    \subfigure[$\langle k \rangle = 4$.]{\includegraphics[width=0.41\textwidth]{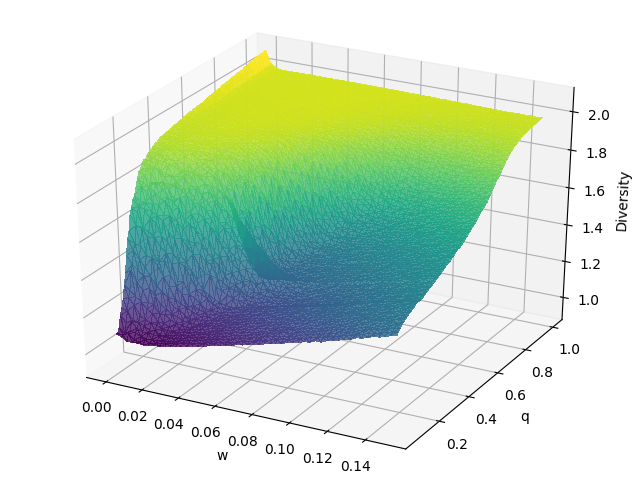}}
    \subfigure[$\langle k \rangle = 8$.]{\includegraphics[width=0.41\textwidth]{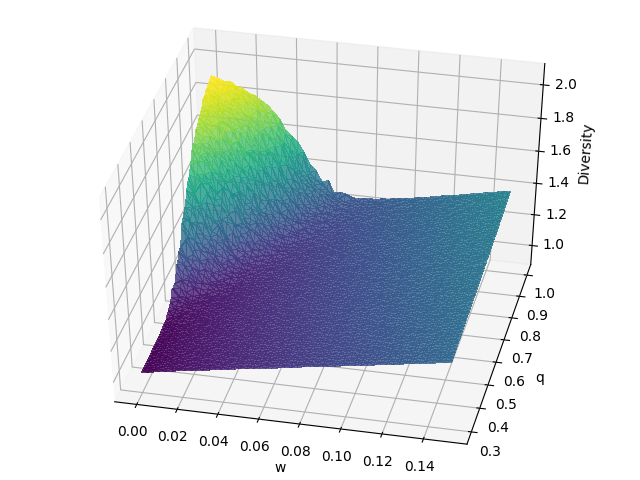}}
    \subfigure[$\langle k \rangle = 12$.]{\includegraphics[width=0.41\textwidth]{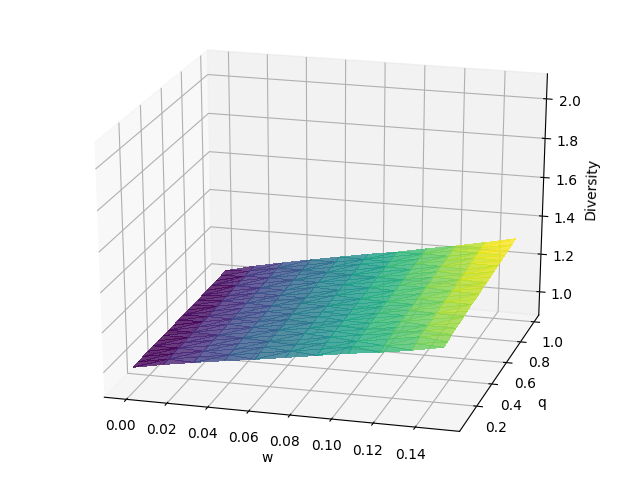}}
  \caption{Mean diversity measurement as a function of $q$ (opinion induced rewiring probability) and $w$ (temperature). The network rewiring probability is $p=0.01$ and the average degree is (a) $\langle k \rangle=4$, (b) $\langle k \rangle=8$ and (c) $\langle k \rangle=12$.}
  \label{fig:diversityXtimeXtemperature}
\end{figure}

\begin{figure}[!htpb]
  \centering
     \includegraphics[width=0.48\textwidth]{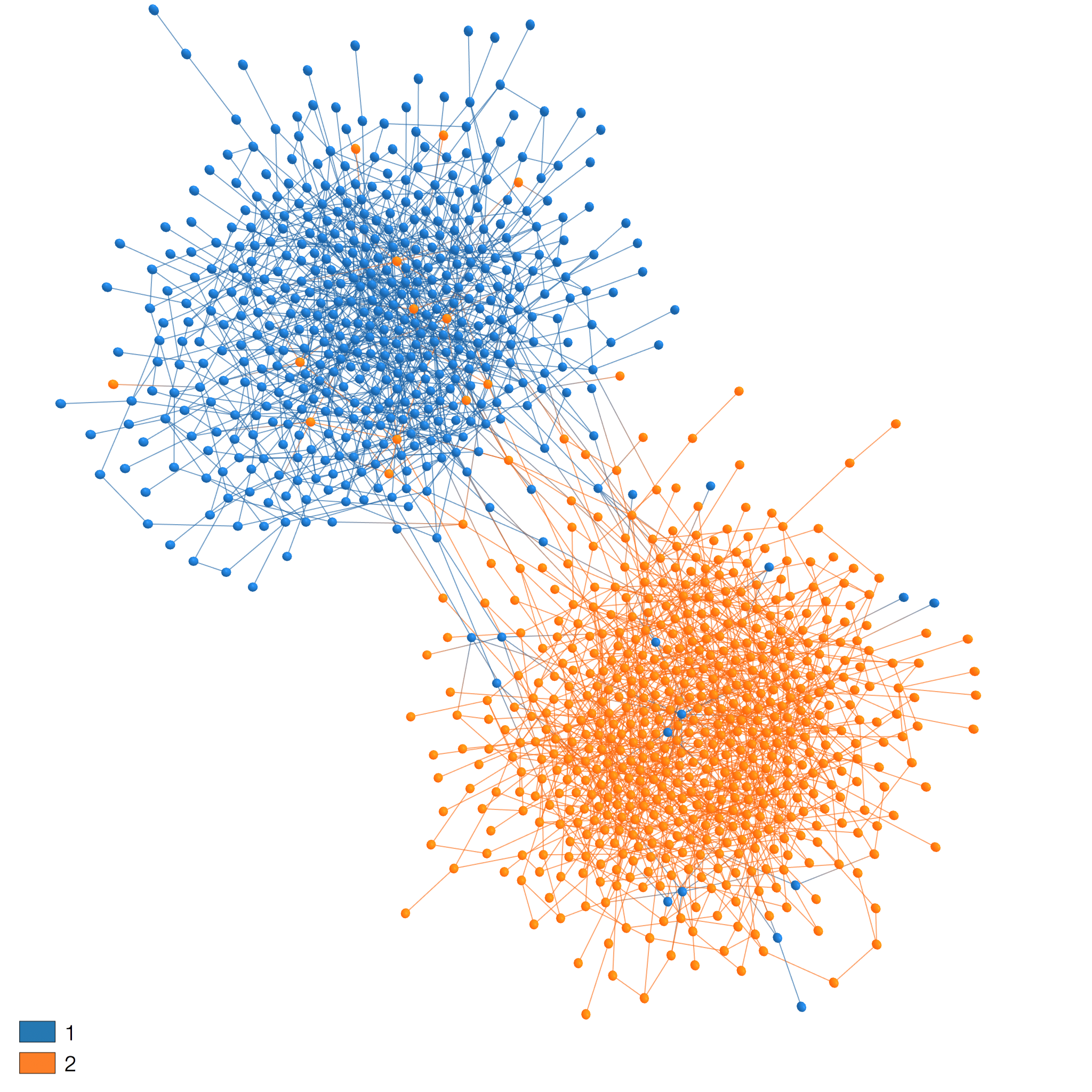}
   \caption{Network visualization by considering 2 opinions and the following parameters: $\langle k \rangle = 4$, $q=0.5$, $w=0.05$, and $p = 0.06$. The colors represent the opinions. This visualization was created using the software implemented in~\citep{silva2016using}.}
  \label{fig:2Opinions}
\end{figure}

In the following, we analyze the relationship among the mean diversity and opinion induced rewiring probability, $q$, and temperature, $w$.
For the sake of simplicity, in the following we fixed the parameter $p=0.01$ as this parameter does not affect significantly the diversity (as shown in  Figure~\ref{fig:diversityXq}).
The obtained results are shown in Figure~\ref{fig:diversityXtimeXtemperature}. Each point in the surface was computed using 500 executions of the dynamics. In Figure~\ref{fig:diversityXtimeXtemperature}(a), we present the results for $\langle k \rangle = 4$, in which there is a well-defined region with diversity approximately equal to 2 (yellow region of the network). A visualization of a typical network obtained for a dynamics having diversity equal to 2 for $w \neq 0$ is shown in Figure~\ref{fig:2Opinions}. Colors represent the opinion of the nodes. The visualization shows that the network becomes divided into two main communities, each community being associated to a different opinion. So, several sets of parameters of the dynamics led to two social bubbles with approximately the same number of agents. When we consider $\langle k \rangle = 8$, this situation also happens, but for a more restricted set of parameters. Contrariwise, for $\langle k \rangle = 12$ there are no social bubbles. As a conclusion, as the average degree increases, the possibility of having a social bubble decreases. Furthermore, the possibility of social bubbles appearing in the dynamics also increases for lower values of $w$ and higher values of $q$.
Interestingly, this type of effect has been studied from several perspectives~\citep{tornberg2018echo,jasny2015empirical}. Some of them analyzed real systems and described the echo characteristics of the chambers~\citep{vicario2019polarization, del2015echo, jasny2018shifting}. In our study, we describe a model that, though simple, can lead to intricate dynamical behavior. 

In another example, we considered a network having 2500 nodes with ten opinions that were randomly assigned as an initial condition. After the transient, the dynamics converge to 10 groups and there are no nodes with the null opinion. Interestingly, because of the temperature ($w = 0.05$), the groups of agents do not generate disconnected components. In other words, some agents are connected to others with different opinions. Additionally, the computed diversity is 9.55, which means that the groups of different opinions have similar sizes. 

\begin{figure}[!htpb]
  \centering
     \includegraphics[width=0.48\textwidth]{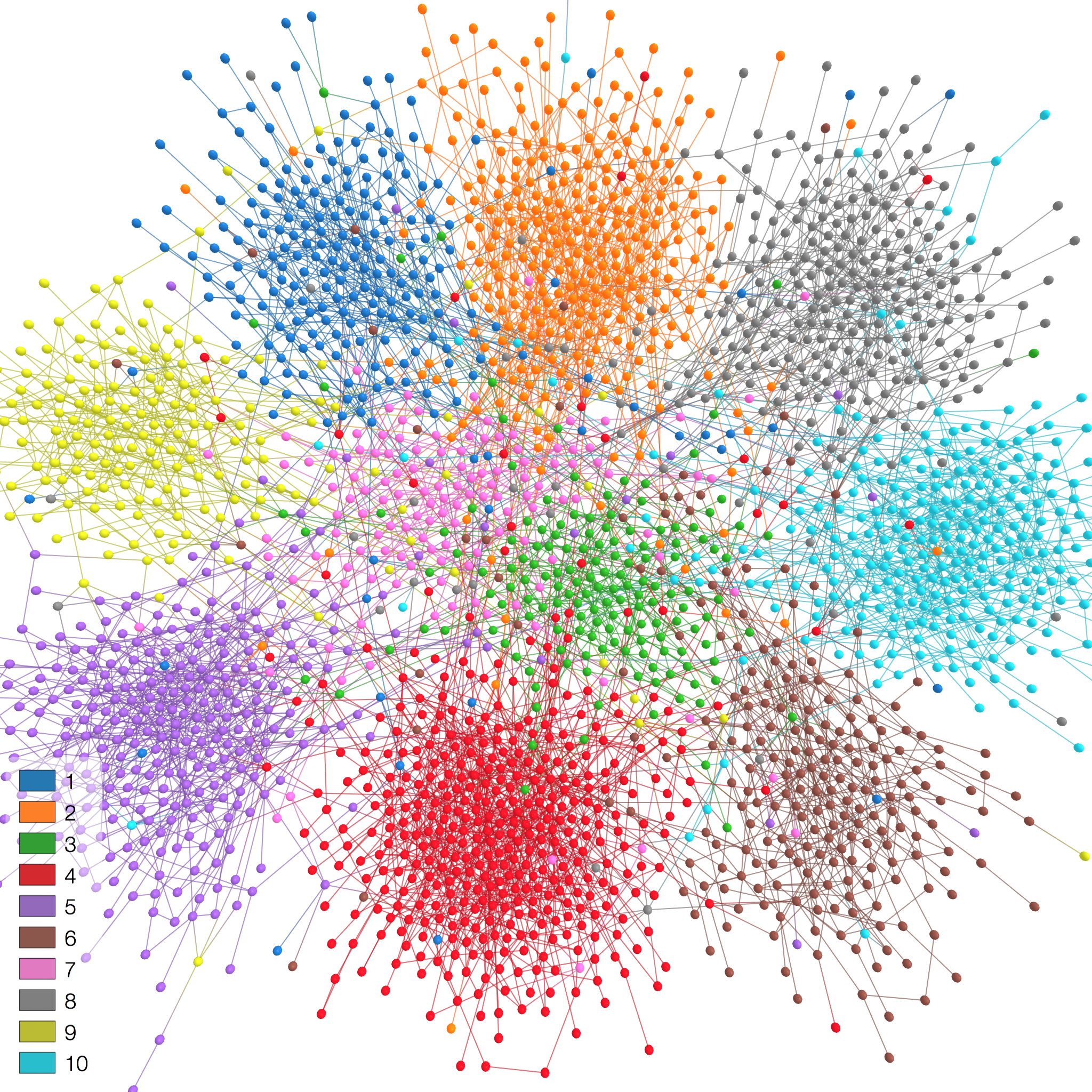}
   \caption{Network visualization by considering ten opinions and the following parameters: $\langle k \rangle = 4$, $q=0.5$, $w=0.05$, and $p = 0.06$. The colors represent the opinions. This visualization was created using the software implemented in~\citep{silva2016using}.}
  \label{fig:10Opinions}
\end{figure}

\subsection{Twin Equilibrium States in the ASM}
\begin{figure}[!htpb]
  \centering
     \includegraphics[width=0.48\textwidth]{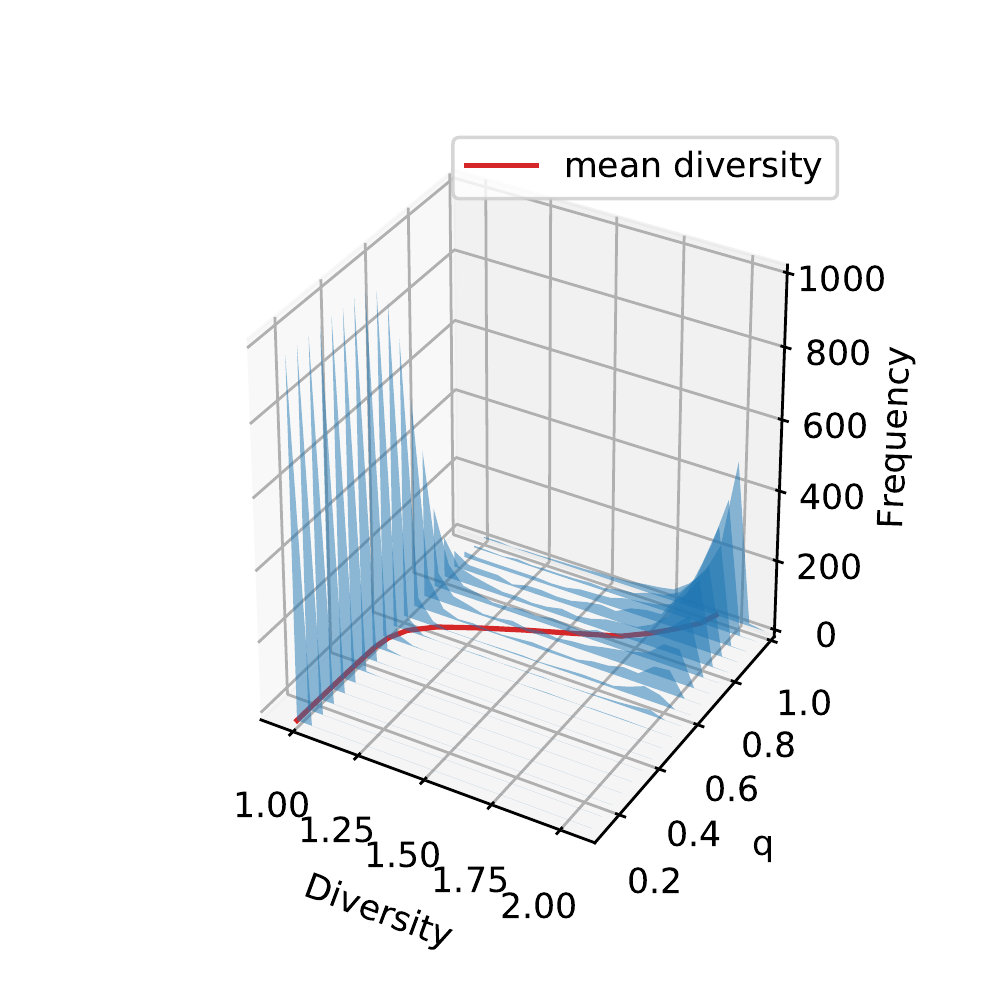}
   \caption{Distributions of diversity values obtained for 1000 realizations of the dynamics for several values of $q$. The remaining parameters of the dynamics are $\langle k \rangle = 8$, $p=0.001$, $w=0$. The red line shows the mean diversity of each histogram.}
  \label{fig:histograms}
\end{figure}

\begin{figure}[!htpb]
  \centering
     \includegraphics[width=0.48\textwidth]{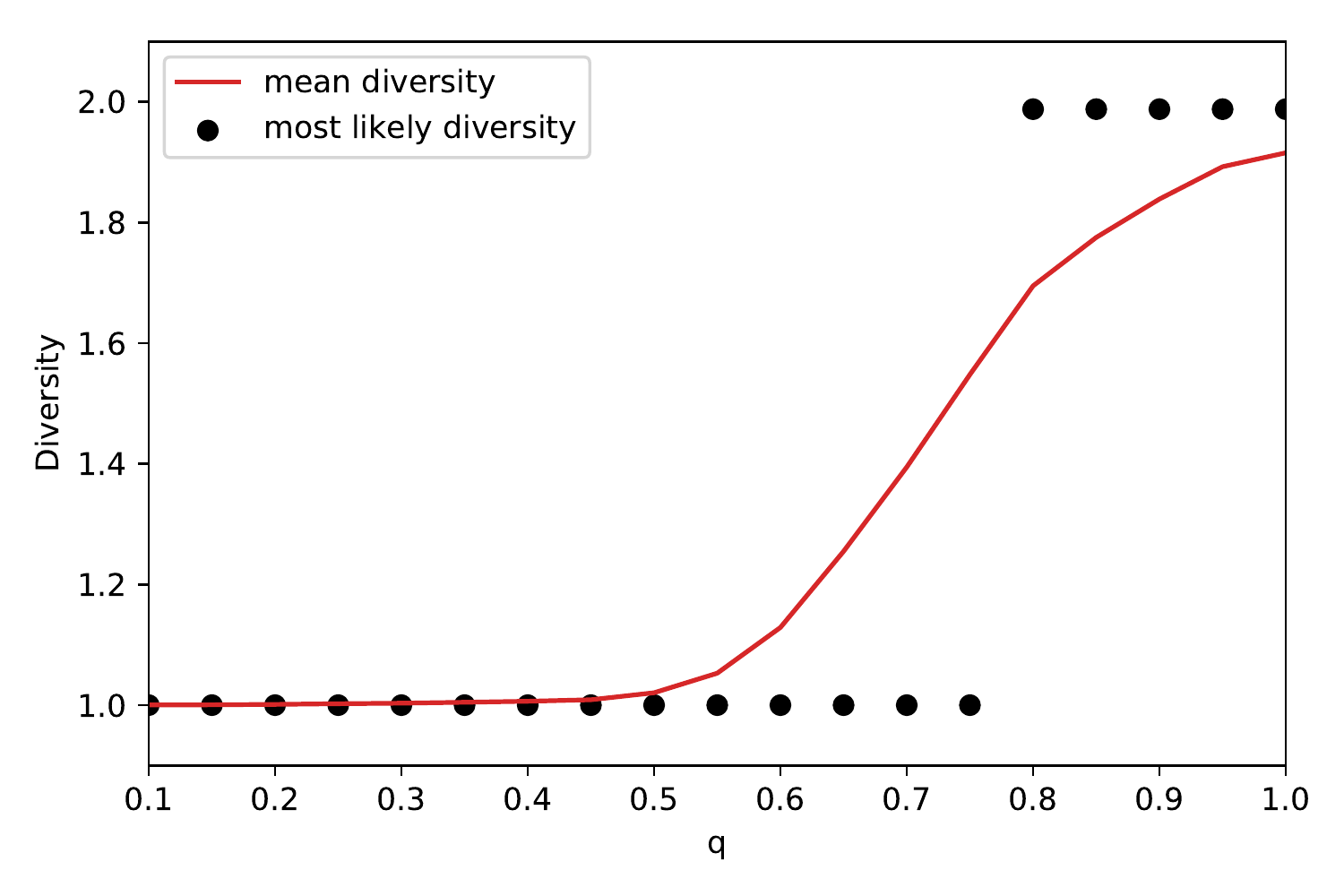}
   \caption{Comparison between the most likely diversity and the mean diversity, according to the histograms presented in Figure~\ref{fig:histograms}.}
  \label{fig:phases}
\end{figure}

In order to better understand the conditions in which social bubbles can happen, we plot histograms of diversity by varying the opinion induced rewiring probability ($q$). The histograms and the respective average diversities are shown in Figure~\ref{fig:histograms}. In this case, we set $w=0$ to analyze only the relationship between diversity and $q$. When $q$ increases, the probability of having higher values of diversity also increases. Interestingly, for all cases, the diversity is usually near 1 or 2. In other words, intermediate diversity values rarely occur, and they can only happen for specific values of $q$. So, its is clear that the average diversity (red line in the plot) cannot correctly describe the behavior of the dynamics in the equilibrium state. We can also analyze the most likely diversity to be attained at the equilibrium state of the dynamics, which is given by the position of the largest peak of each histogram shown in Figure~\ref{fig:histograms}. This information is presented in Figure~\ref{fig:phases} together with the average diversity. The results show that there are two more likely organizations; the first consists of two social bubbles with similar sizes, and the second a single opinion that spans the entire network. 
Therefore, a larger diversity of opinions is usually observed when $q>0.75$. These results suggest that the observed phenomenon could be a phase transition, which is a promising subject for future works.

\section{\label{sec:conc} Conclusions}

The study of human opinion dynamics remains a challenging research activity.  The Sznajd model has been often employed in order to reproduce and predict aspects of opinion changes.  While standard approaches are characterized by a single equilibrium state at zero temperature, it is interesting to consider more flexible behaviors capable of capturing more elaborate aspects of opinion dynamics.  In the present work, we developed an adaptive Sznajd model in which the connections of the network representing the interactions between individuals is allowed to change in order to accommodate neighboring agents to share the same opinion.  More specifically, after changing its opinion, one agent is allowed, with a certain probability, to modify its connections to new neighbors with the same opinion.  

Several interesting results were obtained.  
First, we found that the average degree $\langle k \rangle$ strongly influences the opinion dynamics.  More specifically, the smaller this value, the higher the diversity and the chance of appearing an echo chamber, which is also facilitated by larger values of the parameter $q$ (opinion induced rewiring probability).  The effect of the parameter $w$ is more elaborate in the sense that, when it takes smaller values the chance of obtaining echo chambers is increased.  However, for larger values of $w$, the diversity increases but the probability of having social bubbles tends to become zero.  Overall, it has been observed that small parameter variations can lead to rather distinct opinion dynamics, which corroborates the complexity of such systems and complicates the chances of characterizing and predicting human opinion.  This complexity is further substantiated by the identification that different equilibrium states can be reached for the same parameter configuration.  

Interestingly, our results agree with previous studies involving related dynamics. In~\cite{vazquez2008generic}, the authors show that, depending on the employed parameter, there is the possibility to converge to a unique or two coexisting states. This analysis considered a mean-field approximation. Even though the dynamics are different, the results were found to be qualitatively similar. Furthermore, \cite{holme2006nonequilibrium} proposed a simple dynamics that also give rise to echo chambers, which was analyzed numerically. In comparison with our dynamics, both studies simulate the possibility of having echo chambers. However, with respect to~\citep{holme2006nonequilibrium}, the resultant opinion distributions have not been analogous. We also observe that the formation of social bubbles in the adaptive Sznajd model is in agreement with experimental results~\citep{nikolov2015measuring}.

The reported approach and results pave the way to several future works.  For instance, it would be interesting to consider more opinions, other network types and sizes.  Other possible related research include the characterization of phase transitions and other criteria controlling the topology modifications.

\section*{Acknowledgments}
Alexandre Benatti thanks Coordenação de Aperfeiçoamento de Pessoal de N\'ivel Superior - Brasil (CAPES) - Finance Code 001. Henrique F. de Arruda acknowledges FAPESP (grant no. 2018/10489-0). C\'esar H. Comin thanks FAPESP (grant no. 18/09125-4) for financial support. Luciano da F. Costa thanks CNPq (grant no. 307085/2018-0) and NAP-PRP-USP for sponsorship. This work has been supported also by FAPESP grants 11/50761-2 and 2015/22308-2.

\bibliographystyle{elsarticle-harv}
\bibliography{ref}

\end{document}